\documentclass[aps,prl,twocolumn,
floatfix,superscriptaddress,nofootinbib,
]{revtex4}
\usepackage{amsmath}
\usepackage{amssymb}
\usepackage{epsfig}
\usepackage{graphicx}
\usepackage{pst-coil}

\begin{document}

\title{Split neutrinos - leptogenesis, dark matter and  inflation}  
\author{Anupam Mazumdar } 
\affiliation{Lancaster University, LA1 4YB, United Kingdom, Niels Bohr Institute, Copenhagen University, Denmark}
\author{Stefano~Morisi} 
\affiliation{  AHEP Group, Institut de F\'{\i}sica Corpuscular --
  C.S.I.C./Universitat de Val{\`e}ncia \\
  Edificio Institutos de Paterna, Apt 22085, E--46071 Valencia, Spain}

\begin{abstract}
We propose a simple framework to split neutrinos with a slight departure from tribimaximal mixing - where two of the neutrinos are Majorana type which provide thermal leptogenesis. The Dirac neutrino with a tiny Yukawa coupling explains primordial inflation and the cosmic microwave background radiation, where the inflaton is the {\it gauge invariant} flat direction. The observed baryon asymmetry, and the scale of inflation are intimately tied to the observed reactor angle $\sin\theta_{13}$, which can be further constrained by the LHC and the $0\nu\beta\beta$ experiments.
The model also provides the lightest right handed sneutrino as a part of the inflaton to be the dark matter candidate. \end{abstract}

\maketitle

It is important to connect the origins of inflation, observed neutrino masses, matter-anti-matter/baryon asymmetry, and  dark matter within a falsifiable framework of particle physics beyond the Standard Model (SM), which can be constrained by various low energy observations~\cite{Mazumdar:2011zd}. Since inflation dilutes all matter except the quantum fluctuations which we see in the cosmic microwave background (CMB) radiation, it is important that the inflaton itself cannot be an arbitrary field, its decay must produce the baryons and the dark matter~\cite{Mazumdar:2010sa}.

Furthermore, in order to explain the observed neutrino masses, one must go beyond the SM. In the simplest setting it is possible to augment the SM gauge group by an extra $U(1)_{B-L}$, whose breaking might be responsible for generating the observed neutrino masses. In a supersymmetric (SUSY) setup, this could be realizable within $MSSM\times U(1)_{B-L}$, where (MSSM stands for the minimal supersymmetric SM). 

Gauging $U(1)_{B-L}$ in the SUSY context provides a unique D-flat direction which can be the inflaton candidate as studied previously in Refs.~\cite{hep-ph/0608138,Allahverdi:2007wt}. It was pointed out that a small Dirac Yukawa coupling of order ${\cal O}(10^{-12})$, can actually help  maintaining the flatness of the inflationary potential and provides the right amplitude for the density perturbations, and furthermore - the lightest of the right handed sneutrino (which is now part of the inflaton) can be 
an excellent dark matter candidate~\cite{Allahverdi:2007wt}.

However, neither inflation nor dark matter requires all the three generations to be Dirac in nature. In fact it is quite plausible that two of the other neutrinos could be Majorana type~\cite{Allahverdi:2010us,arXiv:1104.4961}. Since Dirac neutrino does not carry any lepton number, but the Majorana neutrinos do, 
it is now an interesting possibility to realize leptogenesis within this simplest setup. 

In this paper we will demonstrate that it is possible to split neutrinos with one Dirac and two Majorana  types with a non-vanishing reactor angle $\sin\theta_{13}$,  which can explain the baryon asymmetry. The overall neutrino masses are now governed by the Dirac Yukawa and the scale at which the $U(1)_{B-L}$ is broken, therefore achieving inflation, dark matter candidate,  neutrino masses, and baryon asymmetry with a common setup.

Let us first consider for simplicity a single generation of neutrino  with a tiny Dirac Yukawa coupling, $h$.  The superpotential will be given by:
\begin{equation}
W\supset hNH^{u}L
\end{equation}
where $N$, $L$ and $H^{u}$ are superfields containing the right handed neutrino, left-handed 
lepton, and the Higgs which give mass to the up-type quarks, respectively. The  above superpotential generates a renoramlizable potential:
\begin{equation}\label{vpot}
V (\vert \sigma \vert) = \frac{m^2_{\sigma}}{2} \vert \sigma \vert ^2 +
\frac{h^2}{12} \vert \sigma \vert^4 \, - \frac{A h}{6\sqrt{3}}
\vert \sigma \vert^3 \,.
\end{equation}
where $m_{\sigma}^{2}=(m^{2}_{\widetilde N}+m^{2}_{\widetilde L_{1}}+m^{2}_{H^{u}})/3$ is the soft-SUSY mass and it can be
in a wide range, i.e. $m_{\sigma}{\geq \cal O}(500)$~GeV, compatible with the current searches of  SUSY particles at the LHC. The $A$-term is proportional to the inflaton mass $m_{\sigma}$, and the flat direction field $\sigma$ is 
\begin{equation} \label{flat} {\sigma} = ({{\widetilde N} + {H}^{u} + {\widetilde L}_{1})/
\sqrt{3}} , \end{equation}
where ${\widetilde N}$, ${\widetilde L_{1}}$, $H_u$ are the scalar components of
corresponding superfields. Since the right handed sneutrino ${\widetilde N}$ is a
singlet under the SM gauge group, its mass receives the smallest
contribution from quantum corrections due to SM gauge interactions,
and hence it can be set to be the LSP (lightest supersymmetric particle). Therefore the dark matter candidate arises from the right handed sneutrino component of the inflaton.

Inflation happens near the {\it inflection point}~\cite{Allahverdi:2006we,Hotchkiss:2011am},  where $A\sim 4m_{\sigma}$. Near the inflection point it is possible to probe the properties of the 
inflaton~\cite{Allahverdi:2007wt}: $\sigma_0 \approx \sqrt{3}{m_{\sigma}}/{h}=
6 \times 10^{12} ~ m_{\sigma}({0.05 ~
{\rm eV}/ m_\nu})$, and 
$V(\sigma_0) \approx ({m_{\sigma}^4}/{4h^2})=3 \times 10^{24} ~ m^4_{\sigma} ~
({0.05 ~ {\rm eV} / m_\nu})^2 $, 
where $m_\nu$ denotes the neutrino mass which is given by $m_\nu = h
\langle H_u \rangle$, with $\langle H_u \rangle \equiv  v_u \simeq 174$ GeV. 
The largest neutrino mass is $m_\nu \sim \mathcal{O}(1)$ eV\,\cite{Drexlin:2005zt}.
The above potential, see Eq.~(\ref{vpot}), has been studied extensively in order to match the current temperature anisotropy in the 
CMB radiation. It is possible to match the central values of the temperature anisotropy denoted by $\delta_H =
1.91 \times 10^{-5}$ and the spectral tilt: $n_{s}=0.968$, see~\cite{Hotchkiss:2011am}, for a wide range of masses $100~{\rm GeV}\leq  m_{\sigma}\leq 10^9$~GeV and the Yukawa for the Dirac
neutrino in the range: $10^{-12}\leq h \leq 10^{-8}$. The process of reheating and thermalization is quite efficient for a generic flat direction inflaton 
which radially oscillates around its minimum VEV, $\phi=0$, and which carries the SM charges, see for details~\cite{Allahverdi:2011aj}. The temperature at which thermal equilibrium is reached can be estimated to be $T_{\rm R} \leq 10^6$ GeV for $m_\sigma \sim 1$~TeV in our case. 

Scatterings in a thermal bath with the new $U(1)$ gauge interactions also bring the right handed
sneutrino into thermal equilibrium. Note that part of the inflaton, {\it i.e.} its ${\widetilde N}$ component see Eq.~(\ref{flat}), has never
decayed as it is the LSP. Its relic abundance which matches that of the cosmological observations, $\Omega_{CDM}h^2\sim 0.12$
will be then set by thermal freeze-out, which was calculated in Ref.~\cite{Allahverdi:2007wt} for a wide range of the lightest sneutrino mass, i.e. $100~{\rm GeV}\leq m_{\widetilde N}\leq 2000$~GeV.

Although after inflation the reheat temperature is sufficiently high enough to realize the electroweak baryogenesis within MSSM. However, given the current evidence on the Higgs mass searches at the LHC, it is unlikely that the phase transition would be the first order~\cite{Carena:2008vj}. Therefore, one 
would have to rely on other ways of generating the observed matter-anti-matter asymmetry. First of all we would need lepton number carrier, the Dirac neutrinos cannot generate lepton asymmetry.

In our case, the Affleck-Dine baryogenesis would not be realizable even if some of the neutrinos are Majorana. Note that the inflaton is comprised of Eq.~(\ref{flat}), where all the three fields take the same VEVs. Once $NL_{}H^{u}$ takes a large VEV, other directions such as $LH^{u}$, $LLe$, $udd$ can not be lifted at higher VEVs simultaneously. All other directions become massive by virtue of the large inflaton's VEV, see for a review~\cite{Enqvist:2003gh}. The {\it only} plausible scenario would be to realize thermal/non-thermal  leptogenesis. However, this would require at least two of the neutrinos to be of 
Majorana type.

Let us first construct the neutrino masses. It was proposed in Ref.~\cite{Allahverdi:2010us} that two right-handed neutrinos, namely 
$N_e$ and $N_\tau$, could have 
Majorana mass terms, while the $N_\mu$ right handed neutrino has no Majorana mass at the tree-level \footnote{
Majorana neutrino mass term $N_\mu N_\mu$ is forbidden
by means of an Abelian discrete symmetry. },  
and it is coupled to the left-handed neutrino which forms a Dirac mass term.
Since neutrino has a split nature the authors in Ref.~\cite{Allahverdi:2010us} call this
scenario schizophrenic. However the Dirac nature is not protected at higher level, so schizophrenic  
case is equivalent to the quasi-Dirac case \cite{PRINT-82-0694 (SYRACUSE),RL-83-018}.
The overall neutrino masses scale is governed by the Dirac Yukawa $h$. Therefore a  
lower limit for $0\nu\beta\beta$ is obtained in both normal and inverse neutrino mass hierarchy.
Since the second neutrino Majorana mass can be made zero, the limit for $0\nu\beta\beta$ is about a factor of two larger 
than the usual Majorana case and this model can be ruled out very soon by next generation of experiments \cite{0nubbexp}.

The model in Ref.~\cite{Allahverdi:2010us} is based on $S_3$, {\it i.e.} the permutation group of three objects,
see for instance~\cite{Caravaglios:2005gw}. Note that $S_3$ has three irreducible representations, ${\bf 1}$, ${\bf 1'}$ and ${\bf 2}$,
where ${\bf 1'}$ is the antisymmetric singlet.
The relevant product rules are ${\bf 1'} \times {\bf 1'} ={\bf 1}$, and ${\bf 2}\times {\bf 2}={\bf 1}+{\bf 1'}+{\bf 2}$.
In the basis where the generators of $S_3$ are reals, the product
of two doublets, {\it i.e.} $a=(a_1,a_2)$ and $b=(b_1,b_2)$ are given by
\begin{equation}\label{S3p} 
(a_1b_1+a_2 b_2)_{{\bf 1}}+
(a_1b_2-a_2 b_1)_{\bf 1'}+
\left(
\begin{array}{c}
a_1b_2+a_2 b_1\\
a_1b_1-a_2 b_2
\end{array}
\right)_{\bf 2}.
\end{equation}

In order to obtain the neutrinos mass matrix, we extend the SM by introducing 
three right-handed neutrinos: $N_\mu\sim 1$ singlet of $S_3$ and $N_{}=(N_e,N_\tau)\sim 2$ doublet 
of $S_3$. 
As in Ref.\cite{Allahverdi:2010us}, we assume that the combination $L_e,\,L_\mu$ and $L_\tau$ 
\begin{equation}
L_2=\frac{1}{\sqrt{3}}(L_e+L_\mu+L_\tau)\sim {\bf 1}\,,
\end{equation}
transforms as a singlet of $S_3$,
and that 
\begin{equation}\label{L}
L=\left(
\begin{array}{c} 
L_1\\
L_3
\end{array}
\right)=
\left(
\begin{array}{c} 
\frac{1}{\sqrt{2}}(L_\mu-L_\tau)\\
\frac{1}{\sqrt{6}}(-2L_e+L_\mu+L_\tau)
\end{array}
\right)\sim {\bf 2},
\end{equation}
transform as a doublet of $S_3$. Equivalently we assume
that the right-handed charged leptons combination
\begin{equation}
l_2^c=\frac{1}{\sqrt{3}}(l^c_e+l^c_\mu+l^c_\tau)\sim {\bf 1}\,,
\end{equation}
transforms as a singlet of $S_3$, and that 
\begin{equation}\label{L}
l^c=\left(
\begin{array}{c} 
l^c_1\\
l_3^c
\end{array}
\right)=
\left(
\begin{array}{c} 
\frac{1}{\sqrt{2}}(l^c_\mu-l^c_\tau)\\
\frac{1}{\sqrt{6}}(-2l^c_e+l^c_\mu+l^c_\tau)
\end{array}
\right)\sim {\bf 2}.
\end{equation}
We assume two Abelian symmetries $Z_2\times Z_2'$ under which $L_2\sim (+,+)$,
$L \sim (-,+)$, $l_2^c\sim (+,+)$, $l^c\sim (-,-)$, $N_\mu\sim (+,+)$ and $N \sim (-,+)$.
In the scalar sector we assume three sets of $SU_L(2)$ Higgs doublets 
 $H^{u,d}$, $\varphi^{u,d}$ and $\phi^{u,d}$.
These three sets are distinguished by means of $Z_2\times Z_2'$ under which
they transform as $H^{u,d}\sim (+,+)$, $\varphi^{u,d}\sim (-,-)$ and $\phi^{u,d}\sim (+,-)$ respectively. 
The matter content of our model is summarized in table\,\ref{tab1}.

\begin{table}[h!]
\begin{center}
\begin{tabular}{|c|c|c|c|c|c|c||c|c|c|c|}
\hline
 &$\,L_2\,$ & $\,L_{}\,$ & $\,l_2^c\,$ & $\,l_{}^c\,$ & $\,N_\mu\,$ &$\,N_{}\,$ & $\,H_i^{u,d}\,$ & $\,\varphi^{u,d}\,$ & $\,\phi_i^{u,d}\,$ & $\,\xi\,$\\
\hline
$\,\,S_3\,\,$ & ${\bf 1}$ & ${\bf 2}$ & ${\bf 1}$ & ${\bf 2}$ &${\bf 1}$ & ${\bf 2}$ & ${\bf 1},{\bf 1'},{\bf 2}$ & 
${\bf 2}$ & ${\bf 1},{\bf 2}$ & ${\bf 2}$\\
$Z_2$  & $+$ & $-$ & $+$ & $-$ &$+$ & $-$ & $+$ & $-$ & $+$ & $+$\\
$Z_2'$ & $+$ & $+$ & $+$ & $-$ &$+$ & $+$ & $+$ & $-$ & $-$ & $+$\\
\hline
\end{tabular}\caption{Matter content of the model.}\label{tab1}
\end{center}
\end{table}
 
The scalar Higgs doublets $H_i^{u,d}\equiv \{H_{1},H_{1'},H_{2}\}^{u,d}$ transform as ${\bf 1}$, ${\bf 1'}$ and 
${\bf 2}$ with respect to $S_3$ where $H_{2}^{u,d}= (H_{a}^{u,d},H_{b}^{u,d})$. 
Equivalently the Higgs scalar fields  $\phi_i^{u,d} \equiv \{  \phi_{1}^{u,d},\phi_{2}^{u,d}  \}$ with $\phi_2= (  \phi_{a}^{u,d},\phi_{b}^{u,d} )$
and $\varphi^{u,d}= (  \varphi_{a}^{u,d},\varphi_{b}^{u,d} )$ are doublets of $S_3$.

The superpotentials are given by
\begin{eqnarray}
W_l&=&  y_1^l L_2l_{2}^cH_{1}^{d} + y_2^l L_2(l^c\varphi)_2^{d} +  y_3^l (L_{}l^c)_1\phi_{1}^{d} + y_4^l (L_{}l^c)_2\phi_{2}^{d}, \nonumber\\
&& \\
W_\nu&=& h L_2N_\mu H_{1}^{u}+h_s (L N)_1 H_{1}^{u}+h_a (LN)_{1'} H_{1'}^{u}+\nonumber\\
&&\qquad\qquad +h_2 (LN)_2 H_{2}^{u}.\nonumber
\end{eqnarray}
After electroweak symmetry breaking $S_3$ is completely broken, namely $\langle {H_{a}^0}^\alpha \rangle\ne \langle {H_{b}^0}^\alpha \rangle$, 
$\langle  {\phi_{a}^0}^\alpha \rangle\ne \langle  {\phi_{b}^0}^\alpha\rangle$ and 
$\langle  {\varphi_{a}^0}^\alpha \rangle\ne \langle  {\varphi_{b}^0}^\alpha\rangle$ where $\alpha=u,d$. Under this assumption
it is possible to show that the $M_l\cdot M_l^\dagger$ can be  hierarchical and approximatively
diagonal, where $M_l$ is the charged lepton mass matrix. So the lepton mixing arises mainly
from the neutrino sector.
The Dirac neutrino mass matrix is given by
\begin{eqnarray}\label{mdirac2}
&&\quad m_D=\\
&&\left({\small
\begin{array}{ccc} 
-\frac{2}{\sqrt{6}}&\frac{1}{\sqrt{3}}&0\\
\frac{1}{\sqrt{6}}&\frac{1}{\sqrt{3}}&\frac{1}{\sqrt{2}}\\
\frac{1}{\sqrt{6}}&\frac{1}{\sqrt{3}}&-\frac{1}{\sqrt{2}}
\end{array}}
\right)
\left(
\begin{array}{ccc} 
h_s v+h_2u_{b}&0&0\\
0&h v &0\\
0&0&h_sv-h_2 u_{b}
\end{array}
\right)+\nonumber\\
&&+\left({\small
\begin{array}{ccc} 
0&0&-\frac{2}{\sqrt{6}}\\
\frac{1}{\sqrt{2}}&0&\frac{1}{\sqrt{6}}\\
-\frac{1}{\sqrt{2}}&0&\frac{1}{\sqrt{6}}
\end{array}}
\right)
\left(
\begin{array}{ccc} 
h_av'+h_2 u_{a}&0&0\\
0& 0&0\\
0&0&-h_av'+h_2 u_{a}
\end{array}
\right)\nonumber
\end{eqnarray}
where
$\langle H_{1}^{u0} \rangle=v$, $\langle H_{1'}^{u0} \rangle=v'$
$\langle H_{a}^{u0} \rangle=u_{a}$ and $\langle H_{b}^{u0} \rangle=u_{b}$.
Note that in the limit $v',u_a \to 0$ (or $h_a,h_2\to 0$) the Dirac neutrino mass matrix is diagonalized on the
left by tribimaximal mixing $U_{TB}$\,\cite{hep-ph/0202074}\footnote{Tribimaximal mixing $U_{TB}$
 is given by the first matrix in Eq.\,(\ref{mdirac2}).}. For values of $h_a,h_2\ne 0$ we have deviation from tribimaximal
mixing. In particular we generate a deviation of the reactor angle from zero in agreement
with recent T2K \cite{Abe:2011sj} and Double Chooz \cite{DC} experiments. Apparently the reactor angle is a free parameter
in this model (proportional to $h_a,h_2$), however we will show below that it is related to the baryon asymmetry.

Let us now consider the right-handed Majorana neutrinos mass terms. 
We assume a scalar iso-singlet (so coupled only to right-handed neutrino mass terms) $\xi=(\xi_a,\xi_b)\sim 2$ doublet of $S_3$. The superpotential is given by
\begin{equation}
W_M=M(NN)_1+y_\Delta(NN)_2\xi
\end{equation}
Since the term $N_\mu N_\mu$ is missing  the second neutrino mass state $\nu_2$ does not take a 
Majorana mass at tree level and gives rise to a quasi Dirac neutrino mass. Such a term can be forbidden
by means of Abelian symmetries.
For instance, in the Ref.~\cite{Allahverdi:2010us}, the $N_\mu N_\mu$ term was missing by means of the extra $Z_8$ symmetry
under which $N_\mu\to \omega^6 N_\mu$ and a new scalar isosinglet $X\to \omega X$ where $\omega^8=1$. Then the Dirac
coupling of $N_\mu$ is given by $L_2 N_\mu H X^2/M_{\rm P}^2$ where $M_{\rm P}$ is the Planck scale.

We assume $\langle\xi_a \rangle = 0$ and $\langle\xi_b \rangle\ne 0$, then 
the right-handed neutrino mass $M_R$ is diagonal with 
masses
\begin{equation}
M_{N_e}=M+\Delta,\qquad M_{N_\tau} = M-\Delta\,, 
\end{equation}
where the two independent free parameters are respectively, $M$ the $U(1)_{B-L}$ breaking scale, i.e.
$M\sim 1-2$~TeV, and $\Delta=y_\Delta \langle\xi_b \rangle$. In the limit $\Delta \ll M$ the two massive right-handed neutrinos
have degenerate masses. It would be now desirable to have a mass splitting between $N_{e}$ and $N_{\tau}$, since we would like to create the observed matter-anti-matter asymmetry in the universe.

Light neutrino mass matrix arises from type-I seesaw mechanism~\cite{seesaw}, $m_\nu =-m_D\,M_R^{-1}\,m_D^T$
where $m_D$  is defined in Eq.(\ref{mdirac2}). We assume $u_a=0$ in  Eq.(\ref{mdirac2})
and in order to simplify the notation we observe 
that the VEVs $v$, $v'$ and  $u_b$ can be reabsorbed with a re-defination of the Yukawa couplings
$h_s$, $h_a$ and $h_2$ like $h_\alpha\to v_u h_\alpha/v_\alpha $ where $v_u$ is the standard model Higgs
doublet's VEV.

The light-neutrino mass matrix is not diagonal in the tribimaximal basis $U_{TB}$, and it is given by
\begin{eqnarray}\label{mnur}
&&U^T_{TB}\cdot m_\nu \cdot U_{TB}=\\
&&\nonumber\\
&&\left(
\begin{array}{ccc}
\frac{h_a^2}{M_{N_\tau}}+\frac{y_1^2}{M_{N_e}} & 0 &h_a(\frac{y_1}{M_{N_e}}+\frac{y_2^2}{M_{N_\tau}})\\
0&\frac{h}{v_u}&0\\
h_a(\frac{y_1}{M_{N_e}}+\frac{y_2^2}{M_{N_\tau}}) & 0 & \frac{h_a^2}{M_{N_e}}+\frac{y_2^2}{M_{N_\tau}} 
\end{array}
\right)v_u^2\nonumber
\end{eqnarray}
where $y_1=h_2+h_s$ and $y_2=h_2-h_s$. 
When $h_a=0$ the above matrix is diagonal.
In general the matrix in Eq.(\ref{mnur}) is diagonalized by a rotation in the $1-3$ plane $R_{13}(\theta)$.
The lepton mixing matrix is given by $V_l=U_{TB}\cdot R_{13}(\theta)$ and the reactor neutrino
mixing angle $(V_l)_{13}$ is given by 
\begin{equation}\label{theta13}
\sin\theta_{13}= \sqrt{\frac{2}{3}} \sin\theta \approx h_a\sqrt{\frac{2}{3}}\frac{(M_{N_e} y_2+M_{N_\tau} y_1)}{(M_{N_e} y^{\,2}_2-M_{N_\tau} y^{\,2}_1)}.
\end{equation}
The best fit value \cite{Schwetz:2011zk,Fogli:2011qn} of the reactor neutrino mixing angle is about 
$\sin\theta_{13}\sim \mathcal{O}(0.1)$. For small value of the $\theta$ angle, the eigenvalues of the matrix  in Eq.(\ref{mnur})
are approximatively given by
\begin{eqnarray}
m_{\nu 1}&\approx &\left(\frac{h_a^2}{M_{N_\tau}}+\frac{y_1^2}{M_{N_e}}\right)v_u^2,~~~~~~
m_{\nu 2}= h v_u,~
\nonumber \\
m_{\nu 3}& \approx & \left(\frac{h_a^2}{M_{N_e}}+\frac{y_2^2}{M_{N_\tau}}\right)v_u^2. 
\end{eqnarray}
Form this set of equalities we can see immediately that the absolute scale of the neutrinos is fixed by the parameter $h$
that must be about less then $10^{-12}$ in order to have neutrino mass $\mathcal{O}(0.1)$\,eV. Note that from inflation $h \geq 10^{-12}$ for an electroweak scale soft-SUSY breaking masses, therefore predicting large absolute neutrino 
mass scale in our case.
We can obtain $y_{1}$ and $y_{2}$ from the two neutrinos square mass difference 
$\Delta m^2_{atm}$ and $\Delta m^2_{sol}$, namely
\begin{eqnarray}
y_1^2 &=&-h_a^2 \frac{M_{N_e}}{M_{N_\tau}}+\frac{M_{N_e}}{v_u^2}\sqrt{h^2 v_u^2 -\Delta m_{sol}^2} ,\\
y_2^2 &=&-h_a^2 \frac{M_{N_\tau}}{M_{N_e}}-\frac{M_{N_\tau}}{v_u^2}\sqrt{h^2 v_u^2 +\Delta m_{atm}^2-\Delta m_{sol}^2}.\nonumber 
\end{eqnarray}
The parameter  $h_a$ is also related to the reactor angle as it is clear from Eq.(\ref{theta13}).
In particular we can write $\sin \theta_{13}$ as a function of $M$, $\Delta$, $h$ and $h_a$.

Let us now estimate the required $CP$ asymmetry for thermal leptogenesis. The 
asymmetry is calculated by the interference diagrams between tree-level and one-loop 
diagrams, which give rise to~$\epsilon= \sum \epsilon_{\beta\beta}$,~\cite{Roulet:1997xa}
\begin{equation}
\epsilon =
\frac{ \sum_j \mbox{Im}\left[(m_D^\dagger m_D)_{1j}\right]^2}{8 \pi {(m_D^\dagger m_D)}_{11}} g(x_j)
\approx \frac{h_a^2}{2\pi} \frac{M}{\Delta}\sin^2\alpha 
\end{equation}
where  $\epsilon_{\beta\beta}$ is the asymmetry of 
the decay of the right handed (s)neutrinos into $\beta$-family (s)leptons and Higgs, and $x_j=M_j^2/M_1^2$
and we have used the approximation (in the SUSY case) $g(1+z)\approx 2 z^{-1}$ and $\alpha$
is the phase of $h_a$. 
The baryon asymmetry is given by
\begin{equation}
Y_B={\eta_B}/{\eta_\gamma} \approx (\epsilon \,\eta)/{g_{SM}}
\end{equation}
where $g_{SM} = 118$ and $\eta$ is the efficiency factor $\eta\sim \mbox{Min}(1,m^*/m_{\nu_1})$ 
(see for instance \cite{Giudice:2003jh}) where $m_{\nu_1}$ is the lightest neutrino mass and
$m^* = 256\sqrt{g_{SM}}v^2 /(3M_{P}) = 2.3 \cdot 10^{3}\,$eV. Note that the reheat temperature after inflation is sufficiently high to excite the right handed Majorana (s)neurinos from a thermal bath.

\begin{figure}
\begin{center}
\includegraphics[angle=0,height=6cm,width=0.4\textwidth]{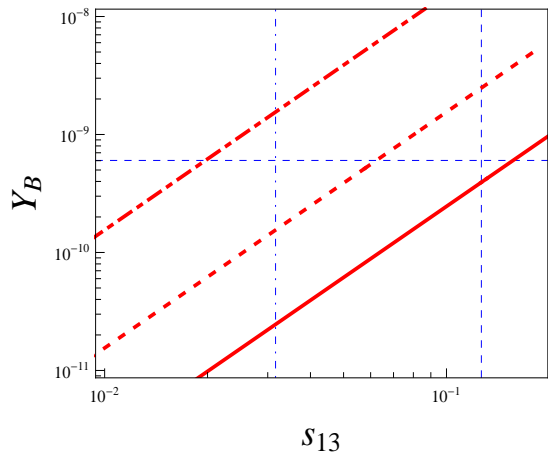}
\caption{Dashed with $M=10^3$GeV, $\Delta=10^{-6}$GeV,
dotdashed with $M=10^4$GeV, $\Delta=10^{-5}$GeV and contineous with $M=10^3$GeV, $\Delta=10^{-5.2}$GeV fixing $h=10^{-12}$.
The horizontal line is the experimental central value of the baryon asymmetry. The two vertical lines are respectively
the $3\sigma$ lower bound and the best fit values for $\sin\theta_{13}$ \cite{Schwetz:2011zk}.}
\label{fig1}
\end{center}
\end{figure}

For fixed values of $M$, $\Delta$ and $\alpha$, the baryon asymmetry is a function of the 
coupling $h_a$. Equivalently, the reactor angle is a function of $h_a$ if we fix $h$, $\Delta m^2_{atm}$ 
and $\Delta m^2_{sol}$, besides  $M$, $\Delta$ and $\alpha$.
In Fig.\,(\ref{fig1}), we show the parametric plot of $Y_B$ versus $s_{13}=\sin\theta_{13}$, varying $h_a$ 
for different choice of the values of  $M$ and $\Delta$ by fixing $\Delta m^2_{sol} , \Delta m^2_{atm}$ at their
best fit values, $\alpha=\pi/2$ (maximal CP violation 
in the lepton sector), and  $h=10^{-12}$ in order to have the neutrino mass scale of about $0.1$\,eV.

To summarize, we have provided a simple realization of split neutrinos where there is one Dirac neutrino 
whose light Yukawa coupling explains the flatness of the inflaton potential, the amplitude of the CMB perturbations, and governs the overall scale 
of neutrino masses through $U(1)_{B-L}$ breaking scale. Besides the Dirac neutrino, there are two Majorana neutrinos with a slight departure from tribimaximal mixing, which explains the reactor angle $\sim\theta_{13}$, and tied intimately to the lepton asymmetry obtained from the decay of the two right handed Majorana (s)neutrinos. This could be a minimal model beyond the SM, where we can explain inflation, dark matter, neutrino masses and the baryon asymmetry, which can be further constrained by the searches of SUSY particles at the LHC, the right handed  sneutrino, essentially the inflaton component as a dark matter candidate, and from the $0\nu\beta\beta$ experiments.

\vskip4.mm
The work of AM was supported by STFC grant ST/J000418/1, and SM was supported by the Spanish MICINN under grants
FPA2008-00319/FPA, FPA2011-22975 and programme Juan de la Cierva, 
MULTIDARK CSD2009-00064, by Prometeo/2009/091 (Generalitat
Valenciana), by the EU ITN UNILHCPITN-GA-2009-237920.

\end{document}